\documentclass[11pt]{article}
\usepackage{blois,epsfig}

\def\Journal#1#2#3#4{{#1} {\bf #2}, #3 (#4)}

\def\PLB{{\em Phys. Lett.}  B}
\def\PRL{\em Phys. Rev. Lett.}
\def\PRD{{\em Phys. Rev.} D}

\def\ApJ{\em Astrophys. J.} 
\def\Nat{\em Nature}
\def\AJ{\em Astron. J.}
\def\MNRAS{\em Mon. Not. R. Astron. Soc.}
\def\IJMPD{{\em Int. J. Mod. Phys.} D}
\def\CQG{\em Class. Quant. Grav.}
\def\RMP{\em Rev. Mod. Phys.}
\def\PAZh{\em Pis'ma A.J.}
\def\SAL{\em Sov. Astron. Lett.}

\def\mco{\multicolumn}

\def\be{\begin{equation}}
\def\ee{\end{equation}}
\def\bea{\begin{eqnarray}}
\def\eea{\end{eqnarray}}

\begin{document}
\vspace*{4cm}
\title{GENERALITY OF INFLATION IN CLOSED FRW MODELS}

\author{ S.A. PAVLUCHENKO }

\address{Sternberg Astronomical Institute, Moscow State University, \\
Moscow 119992, Russia}

\maketitle\abstracts{
We investigate the generality of inflation in closed FRW models for a wide class of quintessence
potentials. It is shown that inflation is not suppressed for most of them for a wide class of
their parameters. This allows us to decide if inflation is common even in case of a closed universe.}

\section{Introduction} 

Recent observations of supernova type Ia (SNIa)~\cite{SN} combined with cosmic microwave background
(CMB) data~\cite{CMB} and data on large scale structure~\cite{LSS} provide us with evidence that
our universe is accelerating now. One can explain this via a presence of a small positive $\Lambda$
term (cosmological constant). Here we consider one kind of dynamical $\Lambda$ term, namely, 
quintessence (see~\cite{alsvar,oth1} for review). It can explain the stage of inflation 
expansion~\cite{infl} and accelerating nowadays; this is the reason for the recent increasing interest in it.
But theories with a scalar field as the source of expansion have a free parameter~-- the potential 
of this scalar field. The aim of this paper is to test some of these potentials that have attracted
attention recently.

To speak about generality of inflation~-- or, in other words, about the probability of inflation for
the model with particular potential one need to introduce the measure on initial conditions space
and, so, parametrize the space of initial conditions. By the term "probability of inflation" we
mean the ratio of the number of solutions experiencing inflation to the number of all possible 
solutions. By {\it the number of solutions} we mean the number one can obtain by using an evently 
distributed net on the space of initial conditions.

\section{Main equations}

The equations describing the evolution of the universe in a closed FRW model are

$$
\frac{m_P^2}{16 \pi}\left(\ddot a+\frac{{\dot a}^2}{2a}+\frac{1}{2a} \right) + \frac{a}{4} \left( 
\frac{{\dot \varphi}^2}{2}- V(\varphi) \right) =0,  
$$
$$
\ddot \varphi + \frac{3 \dot \varphi \dot {\vphantom{\varphi}a} }{a}+\frac{dV(\varphi)}{d\varphi} =0, 
$$
\noindent and the first integral of the system is
$$
\frac{3 m_P^2}{8 \pi}\left(\frac{\dot a^2}{a^2} + \frac{1}{a^2}\right)=\left( V(\varphi)+\frac{{\dot 
\varphi}^2}{2} \right). 
$$

Also we need to introduce the parametrization. We will use {\it trigonometrical (angular)} $(\phi,H)$
and {\it field} $(\varphi,H)$ parametrizations (see~\cite{my4} for details).

Our method is as follows. Starting from the Planck boundary for a given pair of initial conditions 
[$(\phi,H)$ or $(\varphi,H)$] we numerically calculate the further evolution of the universe to 
determine whether universe will experience inflation or not.

\section{Power-law potentials}

Power-law potentials are potentials like 

$$
V(\varphi) = \frac{\lambda \varphi^n}{n}, \quad n \ge 2.
$$

These potentials are well studied and they lead to chaotic inflation~\cite{chaotic}. They have also 
attracted attention for their scaling properties~\cite{PL}. Regarding the degree of inflationarity
for this potential, for $n=2$ for the angular parametrization we have about $63\%$ inflation and for
the field parametrization about $47\%$. Increasing $n$ will decrease the inflationarity in case of the
field parametrization and will not change it for the angular parametrization. In the angular 
measure $63\%$ of all possible solutions experience inflation for $n=2,4,6,8$, and in the field 
measure we have about $30\%$ for $n=4$, $22\%$ for $n=6$, and $17\%$ for $n=8$ (see Table 1 for 
details; three last rows correspond to the cases of initial positive $\dot\varphi$ and negative 
$\dot\varphi$ for the field measure and last row corresponds to the angular measure). 
The Damour-Mukhanov potential~\cite{DM} behaves like a power-law potential in this sence and the 
probability of inflation is about $63\%$ in case of the angular measure and not less than $47\%$ in
case of the field measure (see Table 2 for details).

\begin{table}
\caption{The dependence of the degree of inflationarity on $\lambda$ for different powers of the 
power-law potential.}
\vspace{0.4cm}
\begin{center}
\renewcommand{\arraystretch}{1.8}
\begin{tabular}{||c||c|c|c||c|c|c||c|c|c||}
\hline
& \mco{3}{|c||}{$n = 4$} & \mco{3}{|c||}{$n = 6$} & \mco{3}{|c||}{$n = 8$} 
 \\
\hline
$\lambda$  & $10^{-4}$ & $10^{-5}$ & $10^{-6}$  & $10^{-8}$ & $10^{-9}$ & 
$10^{-10}$  & $10^{-13}$ & $10^{-14}$ & $10^{-15}$ \\
\hline
$+{\dot \varphi}$  & 32.23 & 29.90 & 30.85  & 24.01 & 23.34 & 23.00  & 17.32 & 
16.93 & 18.27  \\
\hline
$-{\dot \varphi}$  & 28.41 & 27.95 & 29.60  & 20.42 & 20.99 & 21.47  & 14.73 & 
15.07 & 16.83  \\
\hline
ang.  & 63.34 & 63.39 & 63.37  & 63.37 & 63.37 & 63.37  & 63.32 & 63.37 & 63.37 \\
\hline
\end{tabular}
\end{center}
\end{table}
\normalsize

\section{Inverse power-law potential}

Pioneering studies of inverse power-law potentials have been done by Ratra and Peebles~\cite{RP}
and these potentials are like

$$
V(\phi) = M^{(4 + q)} \varphi^{-q}.
$$

The dependence of the degree of inflationarity on $M$ is plotted in Fig. 1(a). The degree of 
inflationarity is on the $y$ axis and the power $q$ is on the $x$ axis. There are three curves: 
I corresponds to the case $M \sim m_P$, II to $M \sim 0.7 m_P$, and III to $M \sim 0.4 m_P$. Note
that for this case we can use only the angular measure. 

\section{Exponential potential}

Another interesting potential is exponential one~\cite{RP}:

$$
V(\phi) = V_0 \exp ( - \lambda \varphi  ).
$$

\begin{table}[t]
\caption{The dependence of the degree of inflationarity on $q$ (in columns) and $\varphi_0$ (in rows)
for the Damour-Mukhanov potential.}
\vspace{0.4cm}
\begin{center}
\tabcolsep=1.45em
\renewcommand{\arraystretch}{1.5}
\begin{tabular}{|c|c|c|c|c|c|}
\hline

$\varphi_0 \backslash q$ & 0.5 & 0.8 & 1.0 & 1.5 & 1.8 \\
\hline
0.1 & 84.94 & 74.18 & 67.46 & 55.46 & 50.98 \\
\hline
0.3 & 84.93 & 73.66 & 67.46 & 55.46 & 50.98 \\
\hline
0.5 & 84.63 & 73.66 & 67.46 & 55.46 & 50.98 \\
\hline
0.9 & 84.63 & 74.18 & 67.46 & 55.46 & 50.98 \\
\hline
\end{tabular}
\end{center}
\end{table}
\normalsize

Our results are the same for a wide range of $V_0 \sim m_P^4 \div 10^{-10} m_P^4$. So we have only 
one free parameter, $\lambda$, and the results are plotted in Fig. 1(b). And like in the previous
case we can use only the angular measure.

\begin{figure}[h]  
\vspace*{1.25cm}  
\begin{center}
\epsfig{figure=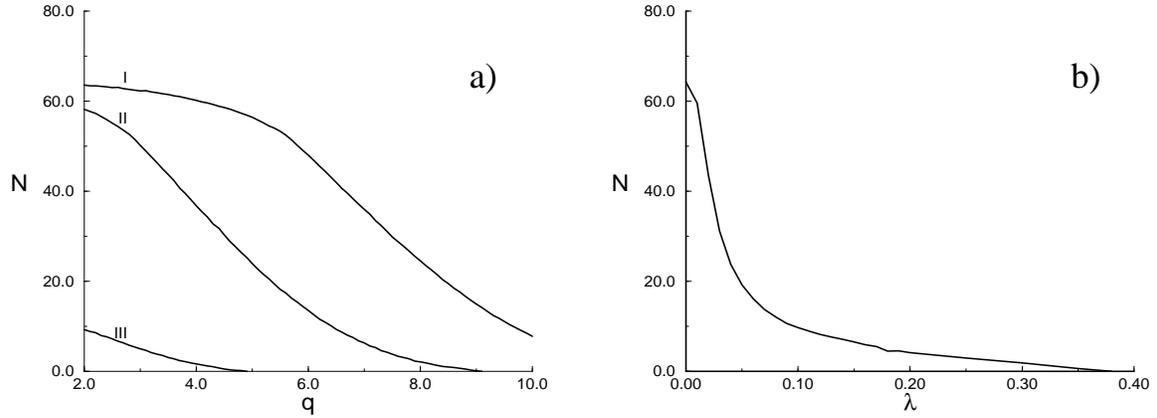,width=16cm}  
\end{center}
\vspace*{0.25cm}  
\caption{The dependence of the degree of inflationarity on the power $q$ for the case of inverse
power-law potential (a) and the same dependence but on $\lambda$ for the case of exponential
potential (b) (see text for details).
} 
\end{figure}

\section{Conclusions} 

In this brief talk we have presented the main results obtained in~\cite{my4}. We have investigated a 
wide
class of quintessence potentials from the point of view of the generality of inflation. And we have
made a weak enough test of them~-- are they able to provide our universe with inflation? And we 
obtained answer yes, closed FRW models with a scalar field with these potentials experience
inflation for a wide range of their parameters. So inflation is general for a wide class of 
cosmological models.
 
\section*{Acknowledgments}
We want to thank the Organizing Committee for hospitality and CNRS and the Organizing Committee for 
financial support. This work was supported by the Russian Ministry of Industry,
Science and Technology through the Leading Scientific School Grant $\#$ 2338.2003.2
We also want to thank N. Yu. Savchenko for 
different help in preparing paper and for useful discussions.

\vfill 
\end{document}